\documentclass{article}
\usepackage[T1]{fontenc} 
\usepackage[utf8]{inputenc} 
\usepackage{ismir,amsmath,cite,url}
\usepackage{graphicx}
\usepackage{color}
\usepackage{amsfonts}
\usepackage{mathtools}
\usepackage[ruled,vlined,linesnumbered]{algorithm2e}
\usepackage{caption}
\usepackage{subcaption}
\usepackage{musicography}

\DeclarePairedDelimiterX{\infdivx}[2]{(}{)}{%
  #1\;\delimsize\|\;#2%
}
\newcommand{\infdiv}{{\mathbb K}{
\mathbb L}\infdivx}

\usepackage{lineno}

\title{Domain Adversarial Training on Conditional Variational Auto-Encoder for Controllable Music Generation}

\multauthor
{Jingwei Zhao$^{2, 4}$ \hspace{1cm} Gus Xia$^{3, 5}$ \hspace{1cm} Ye Wang$^{1, 2, 4}$} {
$^1$ School of Computing, NUS \hspace{.3cm} $^2$ Institute of Data Science, NUS \hspace{.3cm} $^3$ Music X Lab, NYU Shanghai \\ $^4$ Integrative Sciences and Engineering Programme, NUS Graduate School \hspace{.3cm}$^5$ MBZUAI\\
{\tt\small \href{mailto:jzhao@u.nus.edu}{jzhao@u.nus.edu}, \href{mailto:gxia@nyu.edu}{gxia@nyu.edu}, \href{mailto:wangye@comp.nus.edu.sg}{wangye@comp.nus.edu.sg}}
}

\def\authorname{J. Zhao, G. Xia, and Y. Wang}

\usepackage[bookmarks=false,pdfauthor={\authorname},pdfsubject={\papersubject},hidelinks]{hyperref}

\begin{document}

\maketitle
\begin{abstract}
The variational auto-encoder has become a leading framework for symbolic music generation, and a popular research direction is to study how to effectively \textit{control} the generation process. A straightforward way is to control a model using different conditions during inference. However, in music practice, conditions are usually sequential (rather than simple categorical labels), involving rich information that overlaps with the learned representation. Consequently, the decoder gets confused about whether to ``listen to'' the latent representation or the condition, and sometimes just ignores the condition. To solve this problem, we leverage \textit{domain adversarial training} to \textit{disentangle} the representation from condition cues for better control. Specifically, we propose a condition corruption objective that uses the representation to denoise a corrupted condition. Minimized by a discriminator and maximized by the VAE encoder, this objective adversarially induces a condition-invariant representation. In this paper, we focus on the task of melody harmonization\footnote{Demos and codes via \href{https://zhaojw1998.github.io/DAT\_CVAE}{https://zhaojw1998.github.io/DAT\_CVAE}.} to illustrate our idea, while our methodology can be generalized to other controllable generative tasks. Demos and experiments show that our methodology facilitates not only condition-invariant representation learning but also higher-quality controllability compared to baselines.
\end{abstract}

\section{Introduction}\label{sec:intro}

In deep music generation, improving \textit{controllability} has been a major challenge that gains increasing research attention \cite{pati2021disentanglement, dai2021controllable, chen2021controllable, jiang2020transformer, xu2019transferring, kim2019neural}. In practice, controllability is typically implemented under a conditional architecture, where the generation process is biased by external condition inputs. For example, EC$^\mathrm{2}$-VAE \cite{yang2019deep} learns a representation $z_x$ of 8-beat melody $x$ while the underlying chords are given as condition $c$. The system is controllable if the generated melody can adapt to variable chords properly. For such representation-learning architectures, however, the decoder tends to find a shortcut from $z_x$ to $x$ without attending to $c$, leading to ``condition collapse''. The reason for this, as we argue, is that $z_x$ is inevitably intertwined with condition $c$ in the representation space, as $c$ is often an innate property of $x$. In the case of EC$^\mathrm{2}$-VAE, the condition of chords is very much implied by the melody. 



\begin{figure}[htb]
  \centering
  \centerline{\includegraphics[width=.7\linewidth]{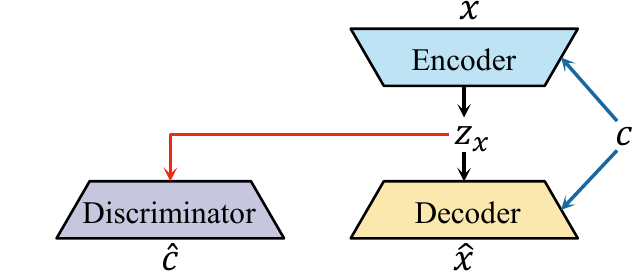}}
\caption{An illustration of domain adversarial training over a conditional generation architecture.}
\label{DAT_illustration}
\end{figure}

To address this problem, the representation $z_x$ must be disentangled from condition $c$. A popular way to achieve this goal is to use an adversarial objective that predicts $c$ from $z_x$, as shown in Figure \ref{DAT_illustration}. On the one hand, this objective is optimized by a discriminator; on the other hand, the encoder is trained to ``fool'' the discriminator by detaching $c$-related cues out of $z_x$. In this way, the decoder cannot find a shortcut in $z_x$ but is forced to seek $c$ to reconstruct $x$. Such a technique stems from \textit{domain adversarial training} (DAT) \cite{ganin2016domain}, where the ``domain'' is interpreted as ``condition'' that controls the generation.

Apparently, DAT can be a powerful tool for controllable music generation. Previous studies \cite{kawai2020attributes, matsuokaattribute} have discussed simple scenarios where the condition is a global label (\textit{e.g.}, note density). In music practice, however, local and sequential conditions \cite{briot2020deep} are more common. In such cases, $c$ may not be fully implied by $x$, so the objective that simply predicts $c$ from $z_x$ does not necessarily hold.




In this paper, we focus on sequential conditions and develop a generalized form of DAT for controllable music generation. We illustrate our methodology with the task of \textit{chord representation learning conditioned on melody}, where $x$ stands for the chord progression, and $c$ is the melody condition. In general, a chord progression can match many melodies, so we cannot directly predict $c$ (melody) from $z_x$ (chord) for the DAT objective. Instead, we leverage $z_x$ to reconstruct $c$ from a corrupted condition $c^*$. We rely on $c^{*}$ to provide the melody context that cannot be hinted by chord $x$; on the other hand, the corrupted information reveals $c$'s harmonic dependency on $x$, which we enforce the discriminator to learn. With proper corruption design, our DAT objective can be generalized to more scenarios with sequential conditions.

A well-trained model with good controllability can help us harmonize a new melody using the representation (style) of an existing chord progression.  Experiments show that our model performs an excellent disentanglement of data representation from the condition, and the controllability outperforms the baselines. In summary, our contributions in this paper are as follows:
\begin{itemize}
    \setlength\itemsep{0em}
    \item \textbf{A general approach to controllability}: Based on a novel adversarial objective with condition corruption, we generalize domain adversarial training to music generation with sequential conditions;
    \item \textbf{A novel harmonization methodology}: We present a representation learning-based method for melody harmonization. Our current model harmonizes pop and folk melodies with the triad and seventh chords. 
\end{itemize}


 

\begin{figure*}[htb]
  \centering
  \centerline{\includegraphics[width=.9\linewidth]{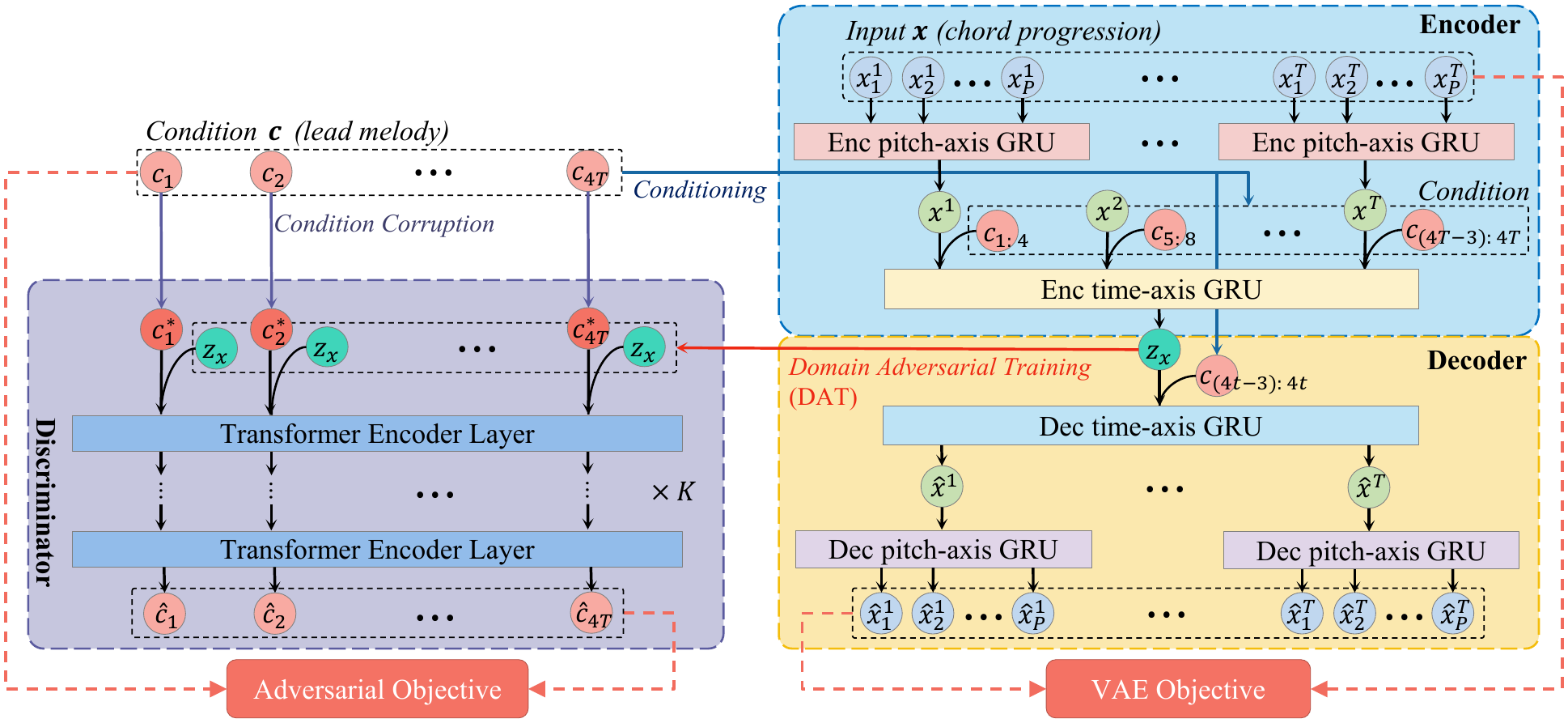}}
\caption{Chord representation learning with adversarial intervention for melody control.}
\label{chdvae}
\end{figure*}

\section{Related Works}

\subsection{Domain Adversarial Training}
Domain adversarial training (DAT) is a representation learning approach initially proposed for domain adaptation tasks \cite{louppe2017learning, wei2020crnn, castellanosunsupervised}. Through an adversarial process as described in Section \ref{sec:intro}, DAT enforces \textit{domain invariance} to data representation so that it can be adapted to different domains flexibly. Such adaptability to new domains is analogous to controllability with new conditions. For generation tasks, DAT has been utilized to learn a condition-invariant data representation. Such invariance enforces the decoder to use condition information for reconstruction \cite{lample2017fader}. During inference, the decoder ``listens to'' new conditions as well and generates new data in a controllable way.

The first attempts that incorporate DAT with generation dealt with facial image generation conditioned on binary attributes (\textit{e.g.}, male or female) \cite{li2016deep, lample2017fader}. Such conditions cannot be explicitly supervised because we cannot find any pair of images that represents the same person both male and female. Fortunately, DAT enforces attribute invariance at encoding and learns attribute dependency at decoding, thus circumventing this problem. Recently, DAT has been extended to symbolic music generation conditioned on various attributes. Kawai \textit{et al.} adopts DAT to a variational auto-encoder (VAE) for melody generation conditioned on statistical attributes (\textit{e.g.}, note density) \cite{ kawai2020attributes}. Later, Matsuoka \textit{et al.} generalizes this methodology to generating polyphonic music with similar conditions \cite{matsuokaattribute}. 

For previous works, the conditions are particularly a global statistical label, which only represents a limited scenario of controllable generation. In our paper, we generalize the usage of DAT to sequential conditions. Conditioned on an 8-bar melody, we aim to learn a pitch-invariant representation of an 8-bar chord progression, which can later be adapted to varied melody conditions and to harmonize them. Our main novelty lies in a special design of the adversarial objective, which is to denoise corruption rather than make full prediction. This technique greatly helps us in dealing with the nuance of sequential conditions.

\subsection{Controllable Music Generation}


Controllable music generation takes various forms in terms of controlling technique and music representation \cite{zhang2020repr}. For controlling technique, controllability can be achieved by sampling, interpolation, conditioning, and more ways \cite{briot2020deep}. For music representation, controls can be performed over statistical music properties (pitch variability, note density, etc.) \cite{kawai2020attributes, matsuokaattribute}, compositional factors (chord progression, texture and rhythmic patterns, etc.) \cite{yang2019deep, wang2020pianotree, wang2020learning, chen2021surprisenet}, high-level semantics (emotion, cultural style, etc.) \cite{zhang2020butter}, and so on. With the development of representation learning, such properties can be abstracted and disentangled for flexible control. 

In this paper, we are interested in chord representation learning conditioned on melodies, which falls into the category of controlling compositional factors via conditioning. Various conditional architectures, such as conditional VAE (C-VAE) \cite{sohn2015learning}, have been applied for similar purposes \cite{yang2019deep, wang2020learning, wang2020pianotree, zhang2020butter, chen2021surprisenet}. However, as the condition is often easily implied by the representation, the decoder tends to skip the condition, and simply reconstruct the data for whatever conditions. To eradicate this problem, we introduce domain adversarial training and generalize it to sequential conditions (in our case, an 8-bar lead melody). Our model learns a pitch-invariant chord representation so that we can generate chord progressions harmoniously conditioned on varied melodies. Such control over compositional factors is common to broader music generation scenarios, and our methodology is generally applicable as well.








\section{Methodology}\label{sec: method}

In this section, we introduce our methodology with domain adversarial training on learning chord representation conditioned on the melody. An overview of our model is illustrated in Figure \ref{chdvae}. We first describe our data representation and structure in Section \ref{2.1}. Then, we introduce our proposed model in Section \ref{2.2}. Finally, we elaborate on our novel design of condition corruption in Section \ref{2.3}.

\subsection{Date Representation and Structure}\label{2.1}
\subsubsection{Chord Representation}
Our model generates an 8-bar chord progression conditioned on the melody. We quantize the chord progression at 1-beat unit and derive $T$ = 32 timesteps. The maximum note count $P$ for each chord is 4, which means we can flexibly represent any type of triad and seventh chords. Specifically, we treat chord progression as a piece of polyphony and follow \cite{wang2020pianotree} to represent it in both a surface structure (as model input) and a deep structure (for encoding).

The surface structure is a nested array of pitch attributes, denoted by $\{x_p^t | 1\leq t \leq T, 1\leq p \leq P \}$. Concretely, $x_p^t$ is the $p^{\mathrm{th}}$ lowest pitch onset at time step $t$. We represent $x_p^t$ as a 13-D one-hot vector corresponding to 12 pitch classes plus a padding state. For most of our chord progression data, the offset of the last chord is precisely followed by the onset of the next one. Hence we do not explicitly consider the duration attributes.

\begin{figure}[htb]
  \centering
  \centerline{\includegraphics[width=.9\linewidth]{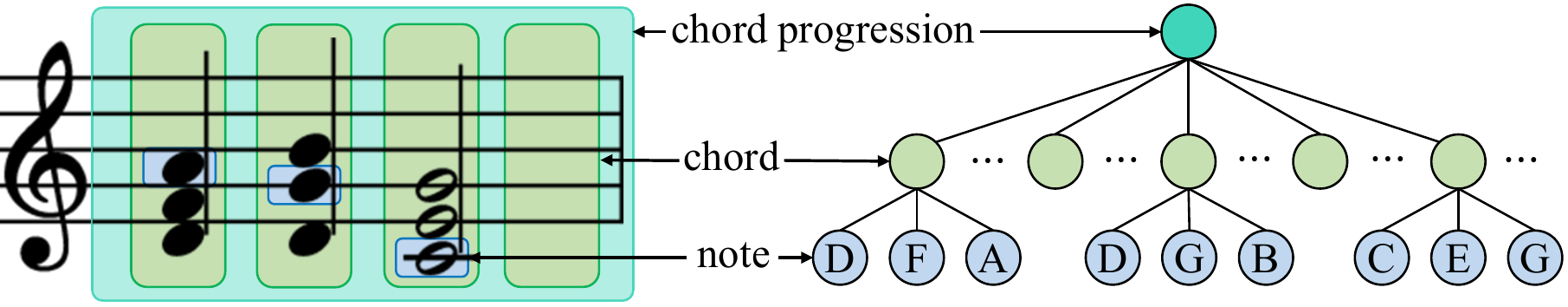}}
\caption{Tree-structure data representation of chord progression, reproduced from \cite{wang2020pianotree} with permission.}
\label{data_representation}
\end{figure}

For the deep structure, we build a syntax tree as in \cite{wang2020pianotree} to reveal the hierarchy from note via chord to chord progression. First, for $1\leq t \leq T, 1\leq p \leq P$, $x_p^t$ itself constitutes the bottom layer of the tree. Then, for $1\leq t \leq T$, we define $x^t$ as the summary of $x_{1\leq p \leq P}^t$, which lies at the middle layer of the tree. Finally, we define $z_x$ as the summary of $x^{1\leq t \leq T}$, which is the root of the tree. Such a deep structure is illustrated in Figure \ref{data_representation}. Conceptually, while $x^t$ is a compact representation of a single chord, $z_x$ represents the complete chord progression.

\subsubsection{Melody Representation}
Our model receives an 8-bar lead melody as the condition. we quantize the melody at $\frac{1}{4}$-beat unit and derive $4T=128$ time steps. Following \cite{yang2019deep}, we represent the melody as a sequence of note onsets plus a hold and a rest state. Each note onset consists of two one-hot vectors each representing 12 pitch classes and 10 octave ranges (registers). In our model, the melody pitch shares the same learnable embedding with the chord pitch. 


\subsection{Proposed Model}\label{2.2}
 Our model applies a similar VAE architecture as PianoTree VAE \cite{wang2020pianotree}, which learns representation for polyphonic music in a hierarchical manner. We use the surface structure of chord progression as the model input. The VAE architecture is built upon the deep tree-like structure.

We first illustrate the vanilla VAE design in the right half of Figure \ref{chdvae}. Let $x$ be the input chord progression and $x_{p}^{t}$ be the $p^{\mathrm{th}}$ lowest pitch onset at time step $t$. The encoder first summarizes $x_{1\leq p \leq P}^{t}$ into an intermediate representation $x^t$ (chord representation) for each time step $t$, and then encodes $x^{1\leq t \leq T}$ to the complete representation $z_x$. The decoder is basically a mirrored version of the encoder. The melody condition $c$, with its every four timesteps summed together, is concatenated to $x^{1\leq t \leq T}$ during encoding and to $z_x$ during decoding. The loss function of our vanilla VAE architecture is:
\begin{equation}\label{melvaeobject}
    \begin{aligned}
    \mathcal{L}(\theta_{\mathrm{enc}}, \theta_{\mathrm{dec}}) &= -\mathbb{E}_Q\left[ \log P_{\theta_\mathrm{dec}}\left(x\mid  z_x, c\right)\right] \\
    &+ \alpha\infdiv{Q_{\theta_\mathrm{enc}}(z_x\mid  x, c)}{\mathcal{N}(\boldsymbol{0},\,\boldsymbol{1})},
\end{aligned}
\end{equation}
where $P_{\theta_\mathrm{dec}}$ and $Q_{\theta_\mathrm{enc}}$ refer to the VAE decoder and encoder. $\theta_\mathrm{dec}$ and $\theta_\mathrm{enc}$ are the learnable parameters. $\alpha$ is a balancing parameter for the regularization of KL loss \cite{higgins2016beta}.

Ideally, $z_x$ should be a \textit{relative} progression representation whose absolute pitch is controlled by melody $c$. However, as the input chord, $x$ already has absolute pitch, this information is preserved in $z_x$ as a redundant melody cue and confuses the decoder from attending to the condition. 

To solve this problem, we assign a {\itshape discriminator} (left in Figure \ref{chdvae}) to the VAE architecture. Instead of predicting $c$ from $z_x$ as conventional DAT objectives do, we bias the discriminator to denoise a \textit{corrupted} melody condition. The corruption is done by transposing the melody to 12 keys with equal chance, which breaks the harmonic relation to the chord. In this way, we learn and extract the chord's dependency on its melody condition. 

Formally, our discriminator leverages $z_x$ to reconstruct melody condition $c$ from a corrupted one $c^*$. Our DAT objective with condition corruption is trained in an adversarial manner. We optimize the discriminator by {\itshape minimizing} the reconstruction loss:
\begin{equation}\label{melad1}
    \mathcal{L}(\theta_{\mathrm{dis}}) = -\mathbb{E}_{Q}\left[ \log R_{\theta_{\mathrm{dis}}}\left(c \mid  z_x, c^{*}\right)\right],
\end{equation}
where $R_{\theta_{\mathrm{dis}}}$ is the discriminator with parameters $\theta_{\mathrm{dis}}$.

%

On the other hand,  we optimize the VAE encoder by {\itshape maximizing} condition reconstruction error:
\begin{equation}\label{melad2}
\begin{aligned}
    \mathcal{L}(\theta_{\mathrm{enc}}\mid  \theta_{\mathrm{dis}}) &= -\mathbb{E}_{Q}\left[ \log R_{\theta_{\mathrm{dis}}}\left(\boldsymbol{1}-c \mid  z_x, c^{*}\right)\right] \\
   &+ \alpha\infdiv{Q_{\theta_{\mathrm{enc}}}(z_x\mid  x, c)}{\mathcal{N}(\boldsymbol{0},\,\boldsymbol{1})},
\end{aligned}
\end{equation}
where $\boldsymbol{1}-c$ is a confusion criterion that encourages the encoder to ``fool'' the discriminator. $\mathcal{L}(\theta_i\mid  \theta_j)$ means we optimize $\theta_i$ while fixing $\theta_j$. The KL loss in Equation \eqref{melad2} and \eqref{melvaeobject} ensures a consistent posterior regularization.

During domain adversarial training, Equation \eqref{melad1} and Equation \eqref{melad2} are iteratively optimized aside from the main VAE objective \eqref{melvaeobject}. In this way, the encoder is explicitly biased to disentangle $z_x$ from $c$. The decoder learns to retrieve missing cues from $c$ to reconstruct $x$, and thus guarantees controllability in the conditional architecture. 

\subsection{Condition Corruption}\label{2.3}
The main novelty of our architecture over previous applications of DAT \cite{lample2017fader, kawai2020attributes, matsuokaattribute} is that we incorporate a corrupted condition term to generalize this method to sequential conditions. The necessity of condition corruption is that, when $c$ is not fully implied by $x$, the conventional DAT objective which predicts $c$ from $z_x$ no longer holds. In our case, $x$ (chord) can be accompanied with various unique $c$ (melodies), and a melody is largely independent of the chord in terms of sequential rhythmic patterns.

Condition corruption aims to reveal the dependency of $c$ on $x$ when a direct predictive inference from $x$ to $c$ cannot be established. The corrupted condition $c^{*}$ serves as a \textit{context} to fill in such prediction gap, and the \textit{dependency} is highlighted when using $z_x$ to denoise $c^{*}$. It may require field knowledge to design a proper corruption method for a specific scenario. Such corruption should keep the context part while blocking the dependency. 

In our case, we corrupt the melody by transposing it to 12 keys with equal probability. The transposed melody $c^*$ keeps the original rhythm and pitch curve shape while distorting the harmonic relation to the chord progression. Here the rhythm and the curve shape are the contexts, and the harmonic relation is the dependency. We compare our corruption method with a corruption-by-masking baseline in Section \ref{object_evaluation} to support the effectiveness of our design.

\section{Experiments}
\subsection{Dataset}
We collect a total of 2K lead sheet pieces (melody with chord progression) for folk and pop songs from Nottingham \cite{nottingham} and POP909 \cite{pop909-ismir2020} datasets. We only keep the pieces with $\frac{\mathrm{2}}{\mathrm{4}}$ and $\frac{\mathrm{4}}{\mathrm{4}}$ meters and slice them into 32-beat snippets at an 8-beat hop size, deriving a total of 35K samples. We quantize chords at $4^\mathrm{th}$ note and melodies at $16^\mathrm{th}$. We randomly split the dataset (at song level) into training (95\%) and validation (5\%) sets. We further augment the training data by transposing each sample to all 12 keys. 

\subsection{Architecture Details}
The VAE framework of our model is consistent with PianoTree VAE \cite{wang2020pianotree}. We implement the encoder with two bi-directional Gated Recurrent Unit (GRU) networks. The pitch-axis GRU and time-axis GRU each has a hidden dimension $d_\mathrm{p, enc}=256$ and $d_\mathrm{t, enc}=512$. The input embedding dimension $d_\mathrm{emb}$ and latent representation dimension $d_\mathrm{z}$ are both set to 128. The decoder mirrors the encoder with uni-directional GRUs, with hidden dimensions $d_\mathrm{t, dec}=1024$ and $d_\mathrm{p, dec}=512$. We set the KL balancing weight $\alpha=0.1$ in Equation \eqref{melvaeobject} and \eqref{melad2}.

We implement the discriminator using BERT \cite{devlin2018bert} with relative positional embedding \cite{shaw2018self, huang2018music, wang2021musebert}, as our condition corruption is conceptually similar to language masking. For our model, we use 4 Transformer encoder layers with 4 heads \cite{vaswani2017attention} and 10\% dropout \cite{hinton2012improving}. The hidden dimensions of self-attention and feed-forward layers are $d_{\mathrm{model}}=256$ and $d_{\mathrm{ff}}=1024$. Our VAE and BERT discriminator each have 12.55M and 3.24M trainable parameters.

\subsection{Training}
Our model is trained using Adam optimizer \cite{kingma2017adam}, with a mini-batch of 256 samples and a learning rate from 1e-3 exponentially decayed to 1e-5. We use teacher forcing \cite{toomarian1992learning} for training the GRU-based decoder, with teacher forcing rate from 0.8 exponentially decayed to 0. We introduce domain adversarial training as an iterative process aside from the main VAE objective, as shown in Algorithm \ref{adversarial}. We set $i=10$, $j=1$, $k=5$, and $l=5$. Our model is trained on a Geforce-2080Ti-12GB GPU. It takes 20 epochs (in around 15 hours) for our model to fully converge. 
\begin{algorithm}[htb]
  \caption{Domain Adversarial Training}\label{adversarial}
  \While{training}
  {
    \For{i iterations}
    {
        Optimize VAE with $\mathcal{L}(\theta_{\mathrm{enc}}, \theta_{\mathrm{dec}}),$
    }
    \For{j iterations}
    {
        \For{k iteration}
        {
            Optimize discriminator with $\mathcal{L}(\theta_{\mathrm{dis}}),$
        }
        \For{l iterations}
        {
            Optimize encoder with $\mathcal{L}(\theta_{\mathrm{enc}}\mid \theta_{\mathrm{dis}}).$
        }
    }
  }
\end{algorithm}

\begin{figure}[htb]
  \centering
  \centerline{\includegraphics[width=.8\linewidth]{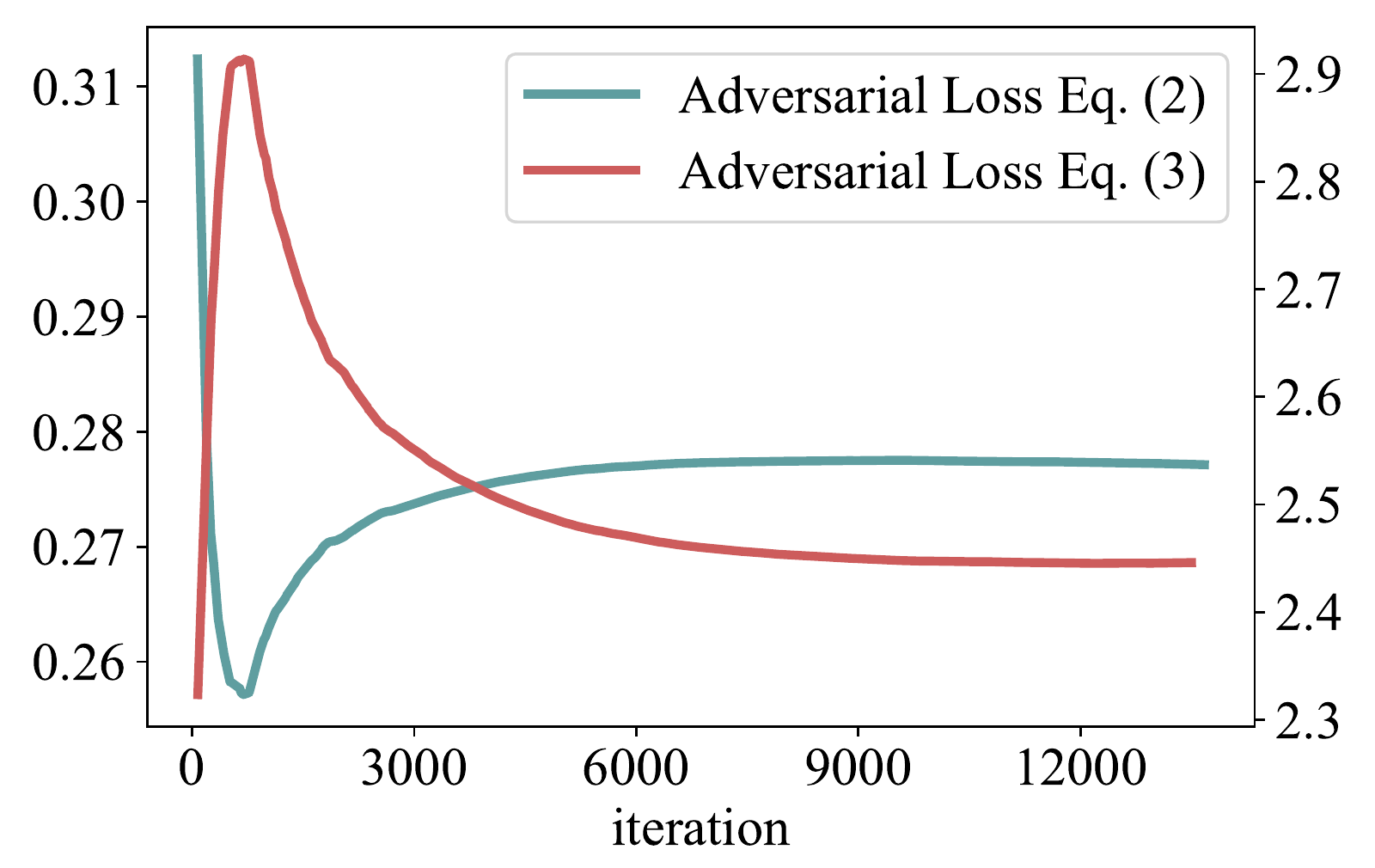}}
\caption{Adversarial loss curves with DAT. Such a trend is driven by the disentanglement of $z_x$ from $c$.}
\label{loss_curve}
\end{figure}

Figure \ref{loss_curve} shows the trends of adversarial loss $\mathcal{L}(\theta_{\mathrm{dis}})$ (in Equation \eqref{melad1}) and $\mathcal{L}(\theta_{\mathrm{enc}}\mid  \theta_{\mathrm{dis}})$ (in Equation \eqref{melad2}). In the early stage, the discriminator learns to reconstruct $c$ based on $z_x$, so the green curve decreases. However, as the adversarial procedure goes on, $z_x$ is gradually disentangled from $c$-related cues. Consequently, the discriminator acquires less and less relevant information to reconstruct $c$ well, and thus the green curve increases. The red curve exhibits an inverse trend, as it is supervised by $\boldsymbol{1}-c$. When each loss curve converges, we interpret it as an equilibrium that indicates a successful disentanglement of chord representation $z_x$ from melody condition $c$.


\subsection{Controllable Generation Results}\label{generation_results}


     
     

\begin{figure}[htb]
     \centering
     \begin{subfigure}[b]{.47\textwidth}
         \centering
         \includegraphics[width=\textwidth]{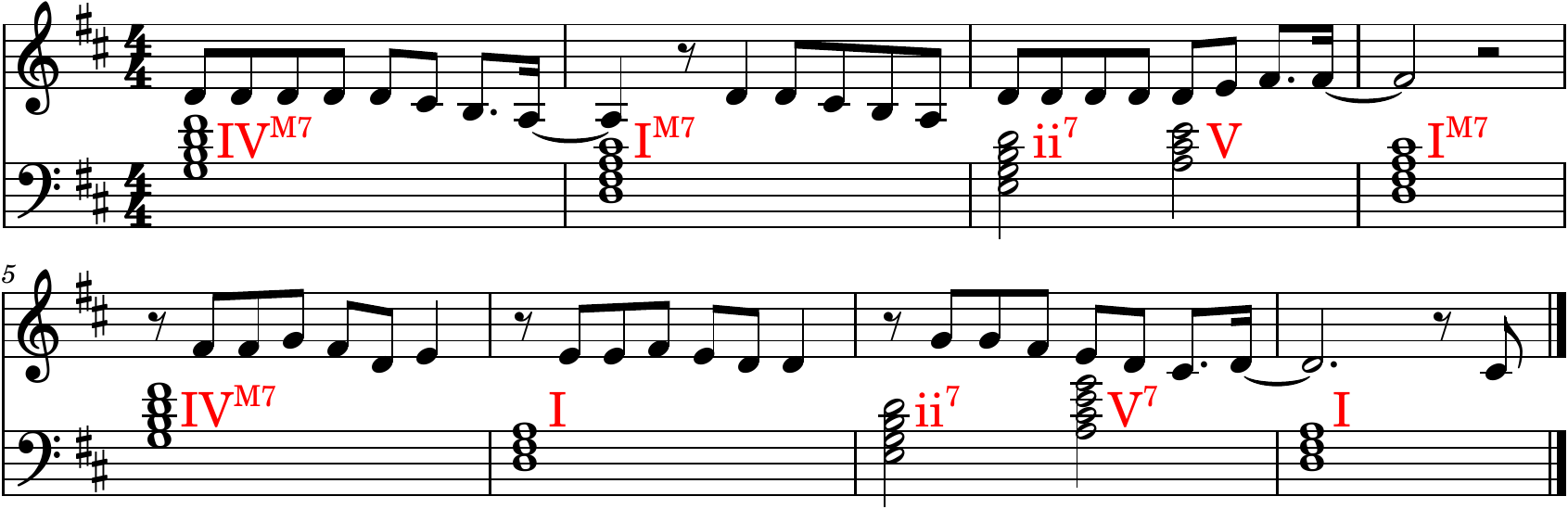}
         \caption{Source A: a D major song accompanied by seventh chords.}
         \label{father}
     \end{subfigure}
     
     \begin{subfigure}[b]{.47\textwidth}
         \centering
         \includegraphics[width=\textwidth]{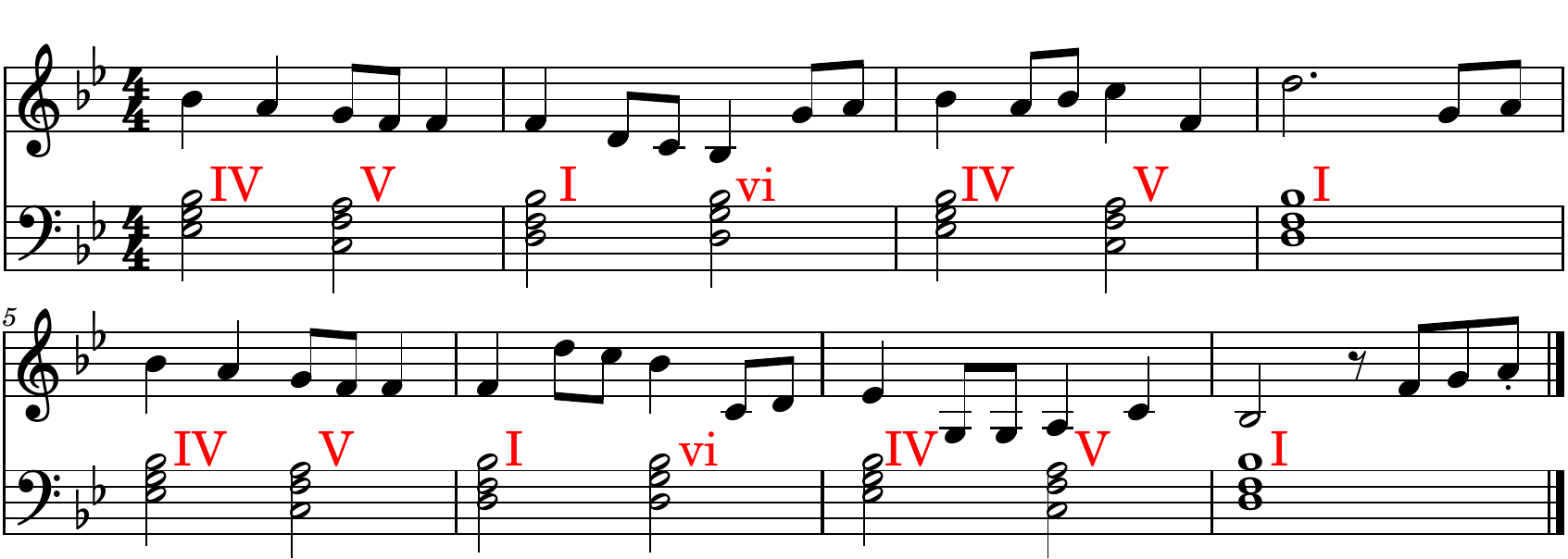}
         \caption{Source B: a B\musFlat{} major song accompanied by triads.}
         \label{mother}
     \end{subfigure}
        \caption{Source lead sheets.}
        \label{father_and_mother}
\end{figure}

Through domain adversarial training, our model gains reliable melody control over chord generation. Our model can harmonize a new melody using the representation of an existing chord progression. We hence develop a novel representation learning-based harmonization methodology. For example, Figure \ref{father_and_mother} presents two source lead sheets selected from our validation dataset. Both source samples are pop song phrases which share similar (but not exactly the same) chord progressions. However, the tonality and chromatic colours of these two pieces are quite different.


\begin{figure}[htb]
     \centering
     \begin{subfigure}[b]{.47\textwidth}
         \centering
         \includegraphics[width=\textwidth]{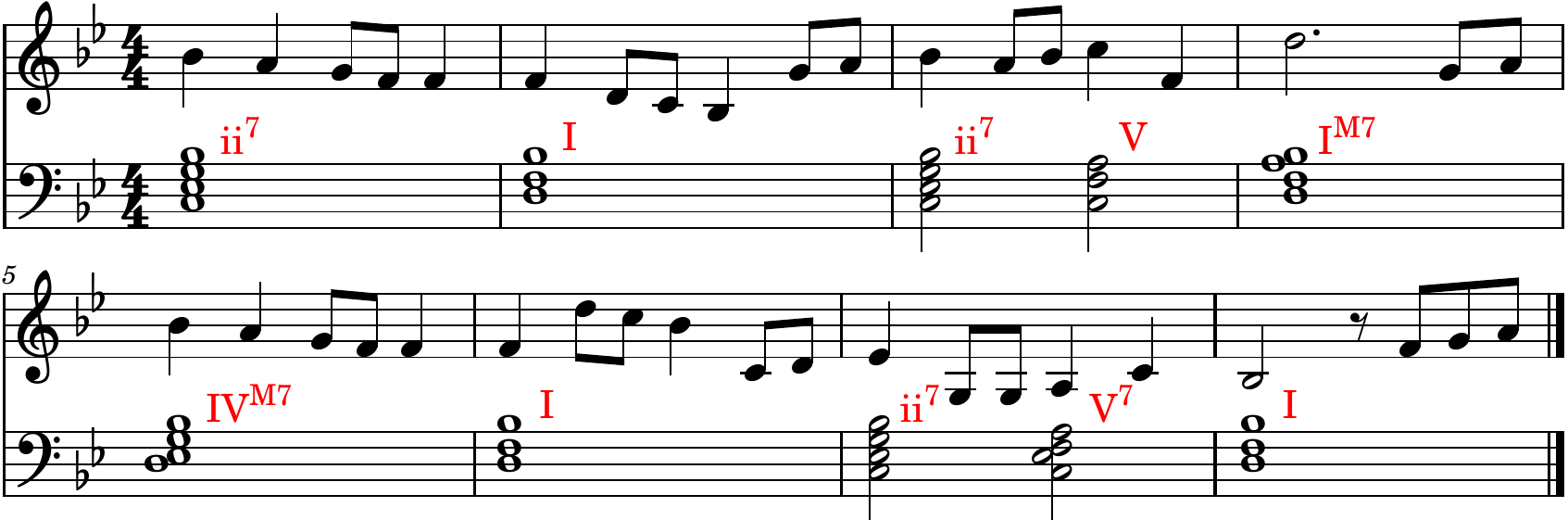}
         \caption{Reconstruction of Chord A conditioned on Melody B.}
         \label{son}
     \end{subfigure}
     
     \begin{subfigure}[b]{.47\textwidth}
         \centering
         \includegraphics[width=\textwidth]{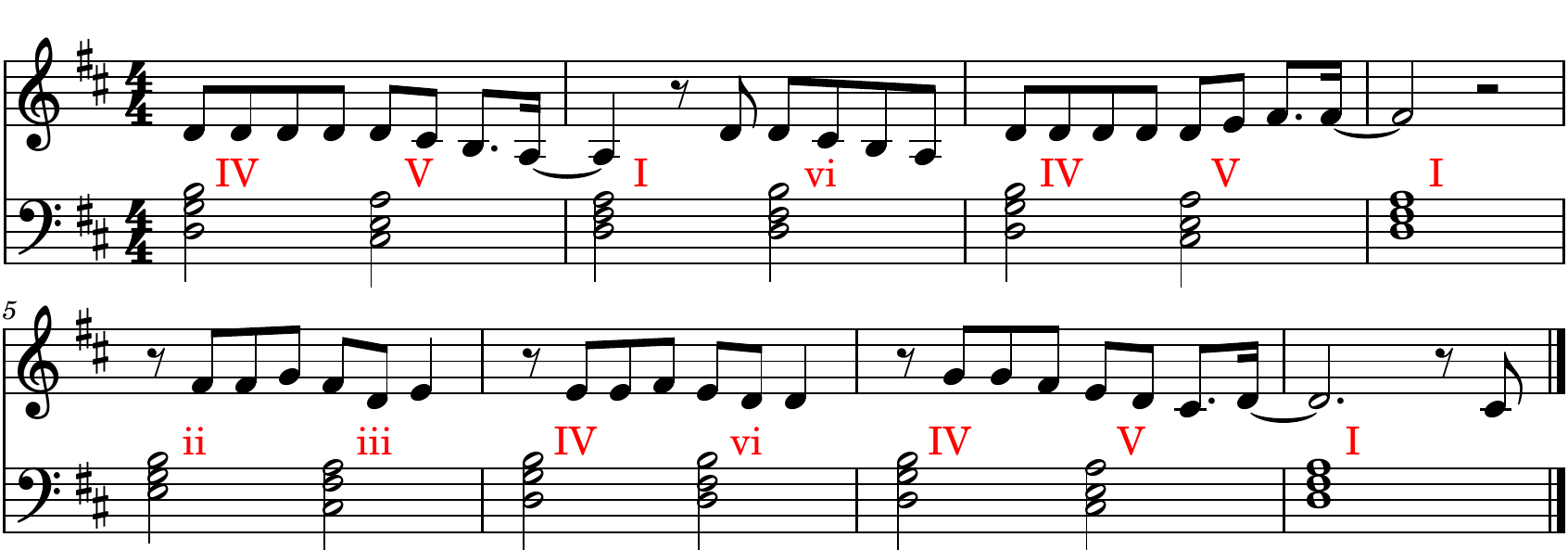}
         \caption{Reconstruction of Chord B conditioned on Melody A.}
         \label{daughter}
     \end{subfigure}
        \caption{Chord generation conditioned on exchanged melody conditions. This process can also be viewed as melody harmonization using exchanged harmonic styles.}
        \label{son_and_daughter}
\end{figure}

Figure \ref{son} is the result where we reconstruct chord A conditioned on melody B, i.e., to harmonize melody B with the harmonic style in A. Here the ``style'' includes tensions with seventh chords and a typical cadence progression of ii-V-I. We see these features properly fitted to melody B in the correct tone. In other words, the generation of chord progression is controlled by the melody. Figure \ref{daughter} is the result where we reconstruct chord B conditioned on melody A. For this case, the original seventh chords in A are replaced by triads with a IV-V-I cadence. These results suggest that our learned chord representation can well discern relative progression and chromatic colour, while our model is controllable in terms of tonality.

\subsection{Subjective Evaluation}\label{subjective_evaluation_Section}
In this section, we evaluate our model's performance on the task of \textit{harmonization}. We first derive the following three baseline models for an ablation study:

\textbf{Non-DAT}: Compared with our model, Non-DAT has the same VAE framework but does not have a discriminator. It does not explicitly try to disentangle $z_x$ from $c$ using domain adversarial training (DAT);

\textbf{Mask-CR}: Mask-CR has the same architecture as our model but uses a different condition corruption technique. Specifically, it applies \textit{masking corruption} (as in \cite{devlin2018bert}) rather than pitch transposition;

\textbf{Non-CR}: Compared with our model, Non-CR uses the conventional DAT objective \textit{without condition corruption}. It predicts $c$ directly from $z_x$ with a GRU discriminator.

To compare our model with the baselines, we survey on rating the harmonization quality of all models. Our survey has 10 groups of harmonization results and each subject is required to listen to 4. In each group, the subjects first listen to an original lead sheet A and a single melody B. Both A and B are 8-bar long (16 seconds) and are randomly selected from different musical pieces from our validation set. As in Section \ref{generation_results}, we harmonize melody B with the harmonic style of A using our model and the baseline models. Subjects are then required to evaluate each version of 
harmonization. The rating is based on a five-point scale from 1 (very poor) to 5 (very high) over three metrics: harmonicity, creativity, and musicality.

\begin{figure}
  \centering
  \centerline{\includegraphics[width=.95\linewidth]{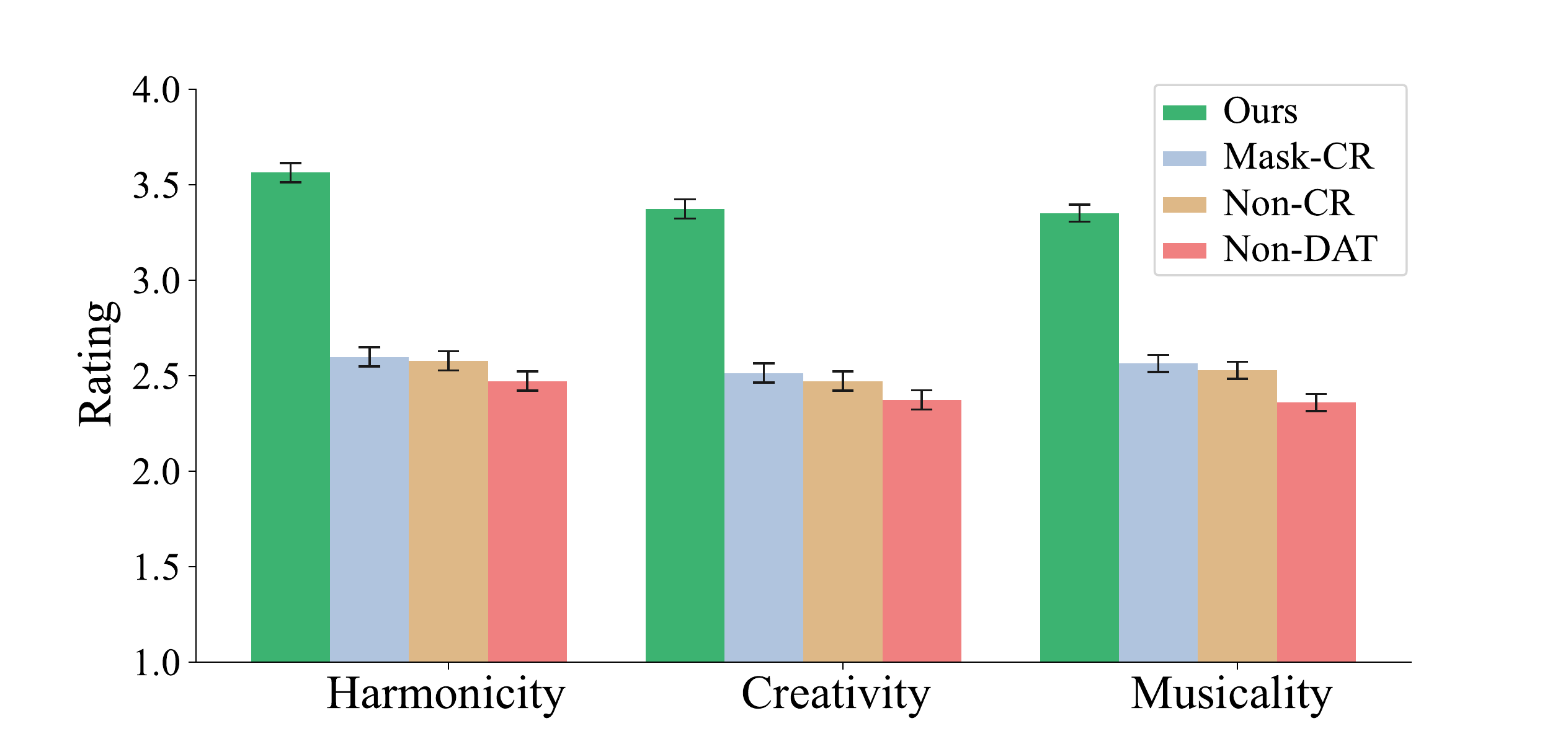}}
\caption{Subjective evaluation on the harmonization performance of our model and baseline models.}
\label{subject_evaluation}
\end{figure}

A total of 38 subjects with diverse music backgrounds participated in our survey and we obtain 142 effective ratings for each metric. As shown in Figure \ref{subject_evaluation}, the height of the bars represents the mean value of the ratings. The error bars represent
the mean square errors (MSEs) computed by within-subject ANOVA \cite{scheffe1999analysis}. We report a significantly better harmonization performance of our model than all three baselines in each metric (p-value $p < 0.05$). Specifically, we note that our model achieves such performance based on a higher degree of representation disentanglement and controllability. We evaluate these methodological aspects with finer objective metrics in the following section.

\subsection{Objective Evaluation}\label{object_evaluation}
In this section, we objectively compare our model with the baselines in terms of \textit{disentanglement} and \textit{controllability}. The baseline models are as defined in Section \ref{subjective_evaluation_Section}.

\subsubsection{Disentanglement}
Our model disentangles chord representation $z_x$ from melody condition $c$. In our case, the melody controls the absolute pitch of the chord progression. A satisfied disentanglement should derive a \textit{pitch-invariant} representation. Following \cite{yang2019deep, wei2021learning}, we develop a similarity criterion to evaluate the performance on disentanglement.

\begin{figure}
    \centering
    \includegraphics[width=.48\textwidth]{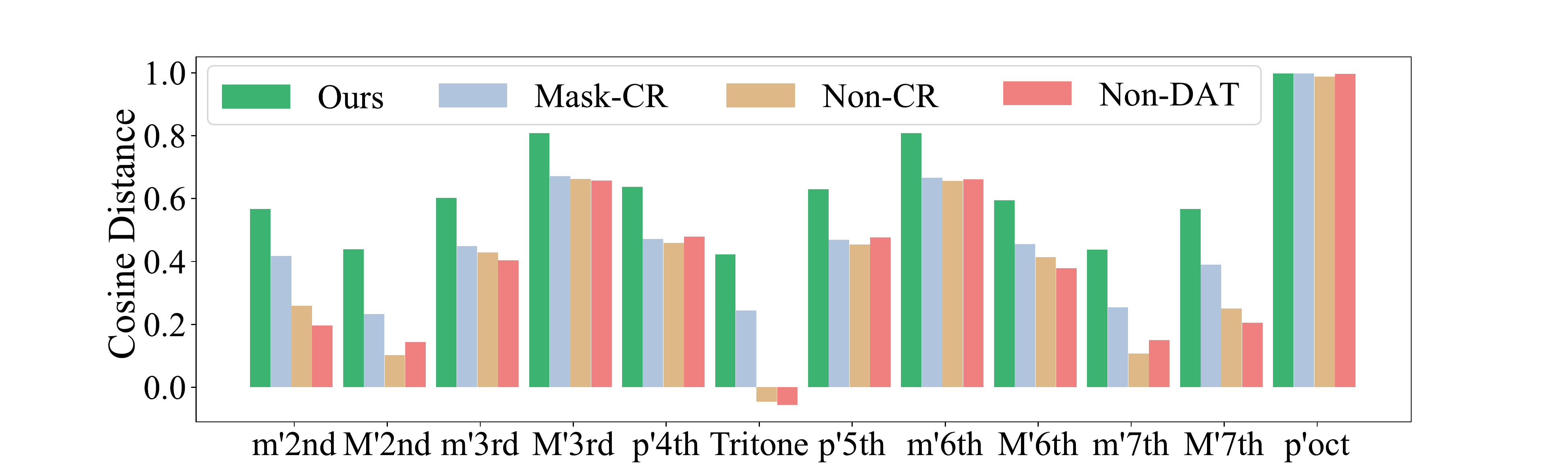}
    \caption{Object evaluation on representation similarity (invariance) against pitch transposition. A higher value denotes better disentanglement.}
    \label{representation_eval}
\end{figure}

Let $\mathrm{T}_i(\cdot)$ be a transposition operator with $i$ semitones. We calculate cosine similarity $\mathrm{cos}(z_x, z_{\mathrm{T}_i(x)}), i =1, 2, \cdots, 12$ for our model and for each baseline. In Figure \ref{representation_eval}, a higher similarity means representation $z_x$ is less affected by the absolute pitch and thus is better disentangled. Our model outperforms all three baselines, including Mask-CR. This finding corroborates that a proper corruption strategy is crucial to applying domain adversarial training to concrete tasks. In our case, masking is not the best way to corrupt, as it is less aware of the harmonization context or dependency discussed in Section \ref{2.3}.

It is also worth noting that the similarity of $z_x$ reflects human pitch perception. For each model, transposing a tritone ($\mathrm{T}_6(\cdot)$) derives the lowest similarity. Figure \ref{representation_eval} shows that $z_{\mathrm{T}_6(x)}$ is literally orthogonal to $z_x$ for Non-DAT and Non-CR. Interestingly, tritone is the most dissonant among all musical intervals in human perception. Such observation indicates that our model learns non-trivial music rules.

\subsubsection{Controllability}
A pitch-invariant representation helps us improve the model controllability by enforcing the decoder to rely on external conditions. In our case of harmonization, a good control generates harmonic chord progression conditioned on the lead melody. Aside from the subjective evaluation in Section \ref{subjective_evaluation_Section}, we introduce \textit{harmony histogram} to objectively interpret the quality of control. Concretely, the harmony histogram is defined as the ratio of within-chord note positions on which the lead melody lies. For tonal music, there should be more root, 3rd, and 5th notes appearing in the melody compared to 7th and higher, so that the music is considered harmonic. 

\begin{figure}
    \centering
    \includegraphics[width=.48\textwidth]{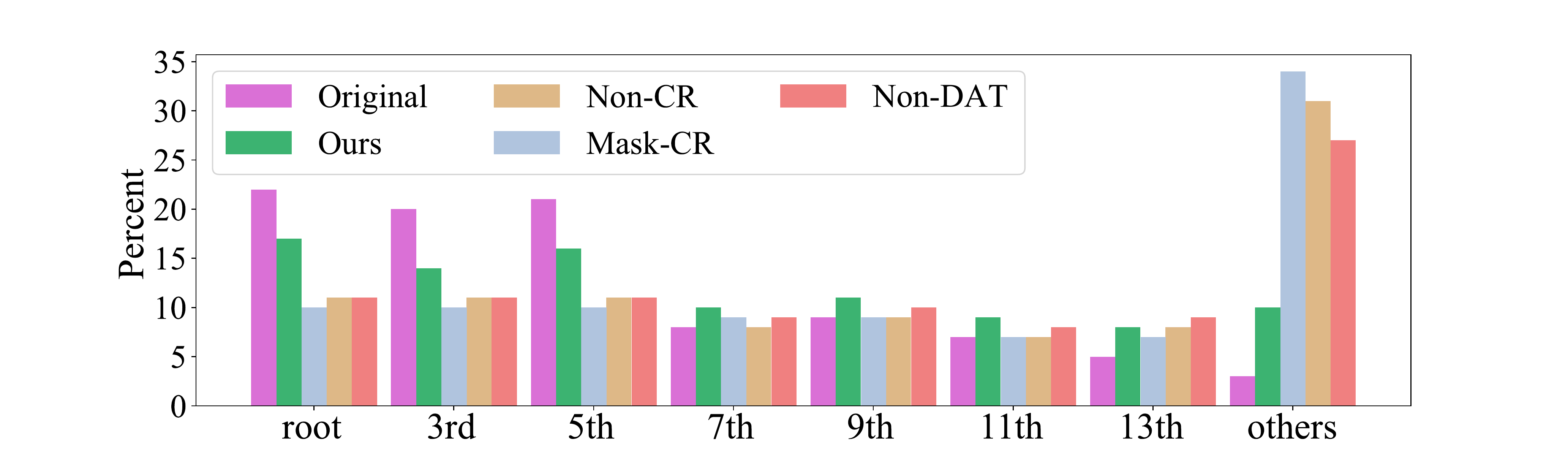}
    \caption{Objective evaluation on harmony histogram upon melody swapping. A higher ratio in root, 3rd, and 5th notes indicates a higher degree of controllability.}
    \label{control_eval}
\end{figure}

In our experiment, we arrange our validation data into random pairs and reconstruct the chord progression with swapped melody conditions. We compare the harmony histogram of generated results from our model and all baselines. Additionally, we compute the histogram for the original (human-composed) data as ground truth. In Figure \ref{control_eval}, we first observe that the histogram distribution has a larger portion in the root, 3rd, and 5th notes for the original data. For the baseline models, over 25\% melody notes are beyond all chord notes and tensions (shown by ``others'' in Figure \ref{control_eval}), which indicates excessive disharmony. Our proposed model, on the other hand, keeps a more consistent pattern with the ground truth. 

\section{Conclusion}
In conclusion, we contribute a generalized form of domain adversarial training for controllable music generation, especially when complex sequential conditions are involved. The main novelty lies in the condition corruption objective, which contextualizes the exact dependency between representation $z_x$ and condition $c$, and therefore assists disentanglement and control. Our method shows excellent performance in chord representation learning, where we learn a pitch-invariant representation conditioned on the melody and develop a novel harmonization strategy. Our improvement in disentanglement and controllability is elaborated with extensive subjective and objective evaluation. With the proposal of our methodology, we hope to bring a new perspective not only to music generation but also to more general scenarios of conditional representation learning.

\bibliography{ISMIRtemplate}

\end{document}